\documentclass{article}

\usepackage{microtype}
\usepackage{graphicx}
\usepackage{subfigure}
\usepackage{booktabs} 

\usepackage{hyperref}

\usepackage[accepted]{icml2024}

\usepackage{amsmath}
\usepackage{amssymb}
\usepackage{mathtools}
\usepackage{amsthm}
\usepackage{mathrsfs}
\usepackage{gensymb}

\newcommand\rbr[1]{\left(#1\right)}

\newcommand\cbr[1]{\left\{#1\right\}}

\usepackage[capitalize,noabbrev]{cleveref}

\theoremstyle{plain}

\theoremstyle{definition}

\theoremstyle{remark}

\newcommand{\ramuno}{i}

\usepackage[textsize=tiny]{todonotes}

\icmltitlerunning{Multi-Region Markovian Gaussian Process}

\begin{document}

\twocolumn[
\icmltitle{Multi-Region Markovian Gaussian Process: An Efficient Method to
Discover Directional Communications Across Multiple Brain Regions}



\icmlsetsymbol{equal}{*}

\begin{icmlauthorlist}
\icmlauthor{Weihan Li}{yyy}
\icmlauthor{Chengrui Li}{yyy}
\icmlauthor{Yule Wang}{yyy}
\icmlauthor{Anqi Wu}{yyy}
\end{icmlauthorlist}

\icmlaffiliation{yyy}{School of Computational Science \& Engineering, Georgia Institute of Technology, Atlanta, USA}

\icmlcorrespondingauthor{Anqi Wu}{anqiwu@gatech.edu}

\icmlkeywords{Machine Learning, ICML}

\vskip 0.3in
]



\printAffiliationsAndNotice{}  

\begin{abstract}
Studying the complex interactions between different brain regions is crucial in neuroscience. Various statistical methods have explored the latent communication across multiple brain regions. Two main categories are the Gaussian Process (GP) and Linear Dynamical System (LDS), each with unique strengths. The GP-based approach effectively discovers latent variables with frequency bands and communication directions. Conversely, the LDS-based approach is computationally efficient but lacks powerful expressiveness in latent representation. In this study, we merge both methodologies by creating an LDS mirroring a multi-output GP, termed Multi-Region Markovian Gaussian Process (MRM-GP). Our work establishes a connection between an LDS and a multi-output GP that explicitly models frequencies and phase delays within the latent space of neural recordings. Consequently, the model achieves a linear inference cost over time points and provides an interpretable low-dimensional representation, revealing communication directions across brain regions and separating oscillatory communications into different frequency bands.
\end{abstract}

\vspace{-0.3in}
\section{Introduction}

The number of simultaneous neural recordings from various brain regions has increased recently. These recordings offer opportunities to explore the mechanisms through which inter-areal communication supports brain function \cite{kohn2020principles}. Brain regions linked to sensory and cognitive functions often display interconnectedness, with signals transmitted bidirectionally and potentially simultaneously \cite{harris2013cortical, miller2018working, wang2024extraction}. However, the high-dimensional neural recordings typically present a complex view of this concurrent communication—for example, neurons may concurrently represent overlapping neural activities within a certain region. Therefore, uncovering the interactions between different brain regions presents a challenging task.

Many statistical methodologies have been employed to address the challenge of understanding communications across multiple brain regions. \citealt{hultman2018brain} examined multi-region local field potential data and identified frequency-based interactions across brain regions using a Gaussian Process Factor Analysis model. Following a similar approach, \citealt{gokcen2022disentangling, gokcen2023uncovering} suggested that latent variables can be divided into across- and within-region components. This model was applied to disentangle the concurrent and bidirectional communications across brain regions with a multi-output Squared Exponential kernel. \citealt{glaser2020recurrent} developed a switching linear dynamic system to uncover low-dimensional interactions among multiple brain regions. This method captured regions responsible for transitioning between latent states by specifying a novel transition rule.

Broadly categorized into Gaussian Process (GP) and Linear Dynamical System (LDS) classes, these methods offer distinct advantages. The GP-based approach, leveraging the robust representational capability of multi-output kernels, performs well in discovering latent variables with crucial information, such as frequencies and directional communications. Conversely, the LDS-based approach, while computationally efficient with a linear cost in time points, lacks the powerful expressiveness of GP in latent representation.

Our goal is to combine the strengths of both methodologies by constructing an LDS that mirrors a GP. Several studies have explored this connection: \citealt{hartikainen2010kalman} established a framework about converting a single-output GP with Matern or Squared Exponential kernel to an LDS, which relied on spectral factorization \cite{sayed2001survey}. Building upon this, \citealt{solin2014explicit} proposed the conversion for single-output periodic kernels, and \citealt{sarkka2013spatiotemporal} extended the single-output conversions to a spatiotemporal GP. However, applying these conversions for GP-based multi-region methods is non-trivial because a gap exists in converting a multi-output GP to an LDS. One approach to bridge this gap is to assume the kernel is separable over spatial and temporal domain \cite{solin2016stochastic}. This allows us to create a closed-form multi-output GP-LDS conversion following the framework proposed in single-output cases.

Consequently, choosing a separable kernel becomes essential in this study. An effective option is the complex-valued multi-region kernel \cite{ulrich2015gp}, specifically designed to facilitate learning latent interactions encompassing frequencies and phase delays across brain regions. However, it is important to note that the connection between an LDS and a GP with a complex-valued multi-output kernel remains unknown.

We introduce the Multi-Region Markovian Gaussian Process (MRM-GP) to model latent representations, where Markovian means the discrete state space representation of a GP. Our work establishes a connection between an LDS and a multi-output GP that explicitly discovers frequency-based latent communications and their directionality via phase delays. By doing so, we can have three advantages: (1)~utilizing the powerful representational capability of kernel functions; (2)~employing the efficient inference algorithm to ensure a linear computational cost over time points; (3)~extending the LDS to incorporate time-varying frequencies and delays by switching states.

We test MRM-GP using multi-region spike trains and local field potential recordings. The model proves its capability to produce understandable low-dimensional representations. These representations illustrate the direction of communication flow among regions and effectively disentangle oscillatory interactions into diverse frequencies.






\section{Background}



We introduce the multi-region kernel, a multi-output kernel for modeling interactions across different brain regions. Then, we demonstrate how a Gaussian Process with this kernel can be employed to model latent communications across regions. It is worth noting that various mapping methods can be used to project latent representations onto neural recordings, and in this case, we opt for Factor Analysis.

\subsection{Multi-Region Kernel}

The complex-valued multi-region kernel proposed in \cite{ulrich2015gp} explicitly models communication frequencies and phase delays within the latent space of neural data: \begin{eqnarray}\label{csm}
\mathbf{K}_{pp'}(\tau)\!\!\!\!&=&\!\!\!\!\sum_{r=1}^{R}a_{p}^{r}a_{p'}^{r}\exp\rbr{
-\frac{1}{2\sigma^2}\tau^2+\ramuno\eta(\tau+\phi_{pp'})}, \\
\!\!\!\!&=&\!\!\!\!\underbrace{\exp\rbr{-\frac{1}{2\sigma^2}\tau^2+\ramuno\eta\tau}}_{\text{temporal}}\underbrace{\sum_{r=1}^{R}a_{p}^{r}a_{p'}^{r}\exp(\ramuno\eta\phi_{pp'})}_{\text{spatial}}.\nonumber
\end{eqnarray} This kernel ensures separability over space and time, where $p$ and $p'$ are two brain regions, $\tau=t-t'$ is the time interval. In the temporal part, $\sigma$ signifies the length scale,  $\ramuno=\sqrt{-1}$ denotes the imagery unit, $\eta$ represents the communication frequency between regions. In the spatial part, $\phi_{pp'}$ represents the phase delay between region $p$ and $p'$, $a_{p}^{r}$ and $a_{p'}^{r}$ are amplitudes, and $R>1$ denotes the rank number ensuring positive definiteness.

The separability is required to establish a connection between the multi-region kernel and a linear dynamic system (LDS). Moreover, the real part of this kernel $\operatorname{Re}[\mathbf{K}_{pp'}(\tau)]=\sum_{r=1}^{R}a_{p}^{r}a_{p'}^{r}\exp(-\frac{1}{2\sigma^2}\tau^2)\cos(\eta(\tau+\phi_{pp'}))$, denoted as the Cross-Spectral Mixture (CSM) kernel \cite{ulrich2015gp}, has been shown to effectively capture frequency-based communications among various brain regions \cite{hultman2018brain}. However, due to CSM's non-separability, we work with the complex-valued kernel as represented in Eq.~\ref{csm} to build an LDS.




\subsection{Multi-Region Gaussian Process Factor Analysis}\label{sec:2.2}

\begin{figure}[t]
  \centering
   \includegraphics[width=\linewidth]{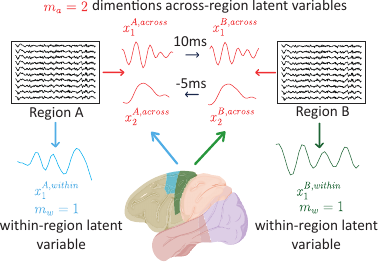}
  \caption{An example of two dimensions across-region latent variables and one dimension within-region latent variable. Brain region A and region B have bidirectional communications within different frequency bands. Each region also has a one-dimensional neural activity unrelated to the other region.}
  \label{fig:1}
\end{figure}

A Gaussian Process Factor Analysis model, utilizing the real part of the multi-region kernel in Eq.~\ref{csm} and named CSM-GPFA, can identify latent variables that capture frequencies and phase delays across brain regions.

Given single region neural recording $y^p \in \mathbb{R}^{n^p \times T}$, $p\in\cbr{1,\dots,P}$ is the brain region index, $n^p$ denotes the number of neurons in region $p$, and $T$ represents time steps. Our goal is to find the $M$ independent low-dimensional variables $x^p \in \mathbb{R}^{M \times T}$ for each region's neural data $y^p$. These variables from $P$ regions together form as $x=[x^1, \dots, x^P]^{\top} \in \mathbb{R}^{MP \times T}$, representing a latent representation for multi-region recordings $y=[y^1, \dots, y^P]^{\top} \in \mathbb{R}^{N\times T}$, where $N=n^1+\dots+n^P$ is the total number of neurons over $P$ regions. Besides, $y$ is a linear mapping of $x$: $y=\mathbf{C}x+d+\epsilon$, where $\mathbf{C}$ is a block diagonal matrix $\mathbf{C}=\text{diag}\{\mathbf{C}^1, \dots, \mathbf{C}^p, \dots, \mathbf{C}^P\} \in \mathbb{R}^{N\times MP}$, $d \in \mathbb{R}^{N \times 1}$ is bias, and $\epsilon\sim \mathcal{N}(0, \mathbf{V})$ is Gaussian noise with $\mathbf{V} \in \mathbb{R}^{N \times N}$.

Meanwhile, a widely used assumption of $x^p$ is to split it into across- and within-region parts \cite{gokcen2022disentangling}: $x^p=[x^{p,a}, x^{p,w}]^{\top}, x^{p,a} \in \mathbb{R}^{m_a \times T}, x^{p,w} \in \mathbb{R}^{m_w \times T}$, where $m_a, m_w$ are the number of dimensions for across- or within-region part and $m_a+m_w=M$. The across-region variables $x^{p,a}$ describe neural activity that is shared across all brain regions, meaning that for the remaining $P-1$ regions, they have the latent variables with the same frequencies and dynamics except phase delays, while the within-region variables $x^{p,w}$ describe the neural activity of region $p$ that is not related to other regions (see Figure~\ref{fig:1}).

Consequently, we model $x^{p,a}$ and $x^{p,w}$ separately with $\mathbf{K}$ in Eq.~\ref{csm}. For region $p$, there are $m_a$ dimensions of across-region variables, and each dimension $x_m^{p,a} \in \mathbb{R}^{T \times 1}, m \in [1, m_a]$ has spatial correlations with the remaining $P-1$ regions. So, the $m^{th}$ dimension across-region variables over $P$ regions: $x^a_m=[x^{1, a}_m, \dots, x^{P, a}_m] \in \mathbb{R}^{P\times T}$ are considered as a group and modeled as the real part of a multi-output Complex Gaussian Process with $\mathbf{K}^{m}$. For within-region variables, each dimension $x^{p,w}_{m} \in \mathbb{R}^{T \times 1}, m \in [1, m_w]$ is independently modeled as the real part of a single-output Complex Gaussian Process with $\mathbf{K}(\tau)^m=\sum_{r=1}^{R}{a_{r}^{m}}^2\exp(-\frac{1}{2{\sigma^m}^2}\tau^2+\ramuno\eta^m\tau)$, where $p=p'$ and $\phi_{pp'}=0$. Furthermore, we also assume independence among different dimensions of across- and within-region variables, implying that different index $m$ refers to distinct kernel parameters.

Unlike the approach in \cite{ulrich2015gp}, which uses a mixture of frequencies, we employ a single frequency in Eq~\ref{csm} to achieve frequency disentanglement. Specifically, each dimension of the across-region latent variable will have a single frequency peak. Consequently, the mixture of frequencies present in the data will be captured by multiple dimensions.

\section{Method}

Modeling latent variables $x^{p}$  with Gaussian Process is inefficient with a $\mathcal{O}(T^3)$ time complexity. So, we want to build Markovian representations of these latent variables, indicating the state space representations of each dimension: across-region $x^{a}_m$ and within-region $x^{p,w}_m$, where every Markovian representation follows a Linear Dynamical System (LDS) \cite{solin2016stochastic}.

Spectral factorization has been used in multi-output GP cases \cite{zhu2023markovian}. However, for the complex-valued multi-output kernel, we need to develop a new spectral factorization-based method to first convert the complex-valued temporal part to an LDS (Section~\ref{sec:3.1}) and then use the kernel's separability to combine the complex-valued spatial part to get the final LDS (Section~\ref{sec:3.2}).

\subsection{Markovian Within-Region Latent Variables}\label{sec:3.1}

The Markovian representation of region $p$'s $m^{th}$ dimension within-region latent variable $x^{p,w}_m \in \mathbb{R}^{T \times 1}$ follows a discrete-time LDS structure: \begin{align}\label{dlds}\begin{split}
f_{m, t}^{p,w}=\mathbf{A}_m^wf_{m, t-1}^{p,w}+q_{t-1}&, \quad q_{t-1} \sim \mathcal{CN}(0, \mathbf{Q}_m^w), \\
\end{split}
\end{align} where $f_{m, t}^{p,w}=[g^{p,w}_{m,t}, \frac{dg^{p,w}_{m,t}}{dt}, \dots, \frac{d^{k-1}g^{p,w}_{m,t}}{dt^{k-1}}]^{T} \in \mathbb{C}^{k\times T}$, denoting the complex-valued dynamics $g^{p,w}_{m,t}$ and its derivatives up to $(k-1)^{th}$ order at time $t$. Especially, within-region latent variable $x^{p,w}_m$ is the real part of $g^{p,w}_{m,t}$.  $\mathbf{A}_m^w \in \mathbb{C}^{k\times k}$ represents the complex-valued transition matrix, and $q_{t-1}$ is the sampling from a complex normal distribution $\mathcal{CN}(\cdot)$ with the complex-valued measurement (Hermitian) matrix $\mathbf{Q}_m^w \in \mathbb{C}^{k\times k}$.

The key question now becomes how to associate single-output $\mathbf{K}(\tau)^m=\sum_{r=1}^{R}{a_{r}^{m}}^2\exp(-\frac{1}{2{\sigma^m}^2}\tau^2+\ramuno\eta^m\tau)$ with $\mathbf{A}_m^w$ and $\mathbf{Q}_m^w$. To achieve this, our approach involves two steps: forming a continuous-time LDS for within-region variables in each region through spectral factorization \cite{kailath2000linear}, and subsequently transforming it into the discrete-time version as specified in Eq.~\ref{dlds}. This linkage is new and differs from previous connections \cite{hartikainen2010kalman, solin2014explicit} as the kernel is situated in the complex domain.

\paragraph{Forming a continuous-time LDS.} Given single-output $\mathbf{K}^m$, the continuous-time LDS we want to form is: \begin{align}\label{clds}\begin{split}
\frac{df(t)^{p, w}_m}{dt}&=\mathbf{F}_m^wf(t)^{p, w}_m+\mathbf{L}u(t),
\end{split}
\end{align} where $f(t)^{p, w}_m=[g(t)^{p,w}_m, \frac{dg(t)^{p,w}_m}{dt}, \dots, \frac{d^{k-1}g(t)^{p,w}_m}{dt ^{k-1}}]^{\top} \in \mathbb{C}^{k\times T}$, denoting the continuous-time version of $g^a_{m,t}$ and its derivatives up to $(k-1)^{th}$ order. $\mathbf{F}_m^w \in \mathbb{C}^{k\times k}$ is a continuous-time transition matrix, $\mathbf{L}=[0,\dots,0,1]^{\top} \in \mathbb{R}^{k\times 1}$ signifies a constant vector, and $u(t)$ denotes a single-dimensional white noise with spectral density $v$. We need to obtain both $\mathbf{F}_m^w$ and $v$ from $\mathbf{K}^m$.

$\mathbf{F}_m^w$ takes a companion form of LDS \cite{grewal2014kalman}: \begin{equation}\label{F}\begin{split}
\mathbf{F}_m^w=\begin{bmatrix}
     0 & 1 &   &     \\
     &  0 & 1 &    \\
     &   & \ddots &  1  \\
    -a_0 &  \dots &  -a_{k-2} &  -a_{k-1}\\
\end{bmatrix}, 
\end{split}\end{equation} where $a_0, \dots, a_{k-1}$ are the coefficients in a stochastic differential equation that is equivalent to Eq.~\ref{clds}: \begin{equation}\label{sde}\begin{split}
\frac{d^{k}g(t)^{p,w}_m}{dt^{k}}+a_{k-1}\frac{d^{k-1}g(t)^{p,w}_m}{dt ^{k-1}}+\dots+a_0g(t)^{p, w}_m=u(t).
\end{split}\end{equation} 

To obtain $\mathbf{F}_m^w$ and $v$, we first apply Fourier transform on both sides of the continuous-time LDS in Eq.~\ref{clds} to achieve a frequency domain representation (see Appendix~\ref{sec:apb} for derivation): \begin{equation}\label{fclds}\begin{split}
S(\omega)=\mathbf{G}(\mathbf{F}_m^w-\ramuno\omega\mathbf{I})^{-1}\mathbf{L}v\mathbf{L}^{\top}(\mathbf{F}_m^w+\ramuno\omega\mathbf{I})^{-T}\mathbf{G}^{\top}
\end{split}\end{equation} where $S(\omega)=\sqrt{2\pi}\sigma^m\exp(-\frac{(\eta^m-\omega)^2}{2\sigma^m})$ is the spectral density of single-output $\mathbf{K}^m$, $\mathbf{G}=[1,0,\dots,0] \in \mathbb{R}^{1\times k}$ represents a constant vector, $\mathbf{I} \in \mathbb{R}^{k\times k}$ denotes an identity matrix, and, notably, $v=\sqrt{2\pi}\sigma^m$. Now, we only need to solve Eq.~\ref{fclds} to obtain the coefficients $a_0, \dots, a_{k-1}$ in $\mathbf{F}_m^w$.

On the left-hand side, $S(\omega)$ follows an exponential family, which is infinitely differentiable. On the right-hand side, however, the finite number coefficients $a_0, \dots, a_{k-1}$ in $\mathbf{F}_m^w$ determine a finite polynomial function: $\mathbf{G}(\mathbf{F}_m^w-\ramuno\omega\mathbf{I})^{-1}\mathbf{L}$. Therefore, we can only construct a finite polynomial approximation of $S(\omega)$. But one observation from Eq.~\ref{fclds} is that $S(\omega)$ can be factorized into two parts, i.e., a complex function multiplying its conjugate.

\begin{figure}[!ht]
  \centering
  \includegraphics[width=1.\linewidth]{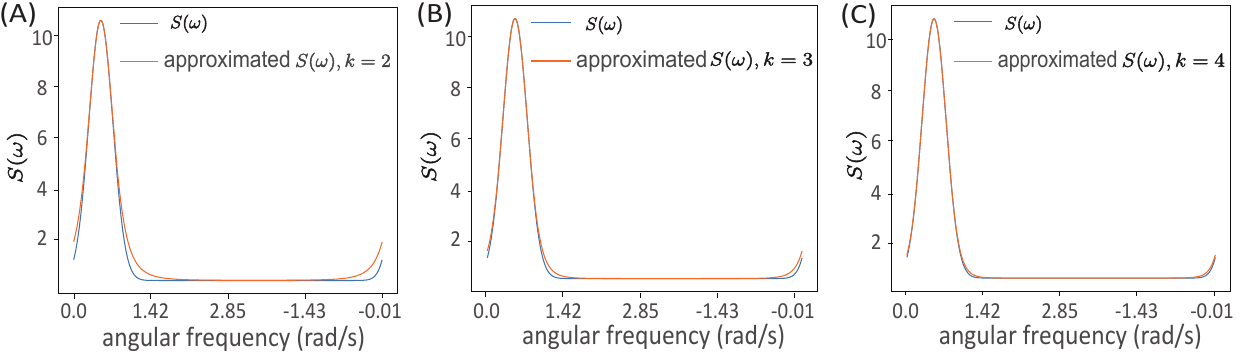}
  \vspace{-3ex}
  \caption{Approximate $S(\omega)$ with $\eta=0.5$ rad/s, $\sigma=4.5$ when varying the number of derivatives $k=2,3,4$. $k=2$ could provide a satisfactory approximation.}
  \label{fig:2}
\end{figure}

Previous established connections between real-valued kernel and LDS assume a symmetric $S(\omega)$, implying using a Taylor expansion to approximate $S(\omega)$ as a polynomial of $\omega$ \cite{hartikainen2010kalman}. However, in the case of complex-valued $\mathbf{K}^m$, $S(\omega)$ is non-symmetric due to frequency $\eta^{m}$, so our solution is to approximate it as a polynomial of $\ramuno\omega$: \begin{eqnarray}\label{taylor}
\frac{1}{S(\omega)}\!\!\!\!&\approx&\!\!\!\!\sqrt{\frac{\sigma^m}{2\pi}}(b_0+b_1\ramuno\omega+b_2(\ramuno\omega)^2+\dots+b_{2k}(\ramuno\omega)^{2k})\nonumber,\\
\!\!\!\!&=&\!\!\!\! T(\ramuno\omega),
\end{eqnarray} where $b_{2k}=1$ if $k$ is even, $b_{2k}=-1$ if $k$ is odd, $b_0, b_2, \dots, b_{2k-2}$ are real numbers, and $b_1, b_3, \dots, b_{2k-1}$ are complex numbers that only have imagery parts. These coefficients' values depend on $\sigma^m$, $\eta^m$, and $k$. Figure~\ref{fig:2}(A-C) shows the approximation of $S(\omega)$ when $k=2,3,4$, demonstrating a reliable approximation even in the case of $k=2$. Appendix~\ref{com_k} shows the effect of $k$ on generated samples.


Now our target becomes solving the following equation for $a_0, \dots, a_{k-1}$: \begin{equation}\label{sf}\begin{split}
T(\ramuno\omega)&=\sqrt{\frac{\sigma^m}{2\pi}}H(\ramuno\omega)H(-\ramuno\omega), \\
H(\ramuno\omega) &= a_0 + a_1\ramuno\omega+\dots+a_{k-1}(\ramuno\omega)^{k-1}+(\ramuno\omega)^{k},
\end{split}\end{equation} where $H(\ramuno\omega)$ is commonly referred to as the transfer function with $a_0, \dots, a_{k-1}$ acting as its coefficients. Its reciprocal $\frac{1}{H(\ramuno\omega)}$ is the function form of $\mathbf{G}(\mathbf{F}_m^w-\ramuno\omega\mathbf{I})^{-1}\mathbf{L}$. Solving Eq.~\ref{sf} is often referred to as spectral factorization, and an advantageous aspect of this factorization is that $a_0, \dots, a_{k-1}$ are a subset of complex-valued roots of $T(\ramuno\omega)$, where they are all situated within the left-half complex plane \cite{kailath2000linear} and can be found by QR algorithm with time complexity $\mathcal{O}(k)$. See Appendix~\ref{sec:apb} for derivation. 


\paragraph{Forming a discrete-time LDS.} Given $\mathbf{F}_m^w$, $\mathbf{L}$ and $v=\sqrt{\frac{2\pi}{\sigma^m}}$ in Eq.~\ref{fclds}, the computation of $\mathbf{A}_m^w$ and $\mathbf{Q}_m^w$ in Eq.~\ref{dlds} are as follows \cite{solin2016stochastic}: \begin{align}\label{Lyapunov}\begin{split}
\frac{d\mathbf{P}_{\infty}}{dt}&=\mathbf{F}_m^w\mathbf{P}_{\infty}+\mathbf{P}_{\infty}{\mathbf{F}_m^w}^H+\mathbf{L}v\mathbf{L}^{\top}=0, \\
\mathbf{A}_m^w&=\text{expm}(\mathbf{F}_m^w\Delta t), \\
\mathbf{Q}_m^w&=\mathbf{P}_{\infty}-\mathbf{A}_m^w\mathbf{P}_{\infty}{\mathbf{A}_m^w}^H,
\end{split}
\end{align} where $v$ is the spectral density of white noise $u(t)$, $\text{expm}(\cdot)$ represents the matrix exponential function, and $\Delta t$ signifies the time interval in discrete-time LDS.

\subsection{Markovian Across-Region Latent Variables}\label{sec:3.2}

Region $p$'s $m^{th}$ dimension across-region latent variable $x^{a}_m \in \mathbb{R}^{P \times T}$ can be expressed as $x^a_m=[x^{1, a}_m, \dots, x^{p, a}_m, \dots, x^{P, a}_m]$. This indicates that they consist of $P$ variables sharing the same $\eta^m$ and $\sigma^m$ for describing temporal features while using phase delays $\{\phi_{pp'}^m\}_{p,p'=1}^P$ to capture cross-spatial differences. It's important to note that the temporal and spatial components are separable in $\mathbf{K}^m$. Consequently, we can initially create a within-region Markovian representation, denoted as $\mathbf{A}_m^w, \mathbf{Q}_m^w$, for the temporal features in each $x^{p, a}_m$. Then, this representation is extended to the across-region Markovian representation for $x^{a}_m$ through the incorporation of phase delays $\{\phi_{pp'}^m\}_{p,p'=1}^P$.

Therefore, the Markovian representation of $m^{th}$ dimension across-region latent variable $x^{a}_m$ is: \begin{align}\label{dlds2}\begin{split}
f_{m, t}^{a}=\mathbf{A}_m^af_{m, t-1}^{a}+q_{t-1}, &\quad q_{t-1} \sim \mathcal{CN}(0, \mathbf{Q}_m^a), \\
\mathbf{A}_m^a=\mathbf{I}\otimes\mathbf{A}_m^w, &\quad \mathbf{Q}_m^a=\mathbf{K}^m_{\text{spatial}}\otimes\mathbf{Q}_m^w,
\end{split}
\end{align} where $f_{m, t}^{a}=[g^a_{m,t}, \frac{dg^{a}_{m,t}}{dt}, \dots, \frac{d^{k-1}g^{a}_{m,t}}{dt^{k-1}}]^{\top} \in \mathbb{C}^{Pk\times T}$, $x^a_{m,t}$ is the real part of $g^a_{m,t}$, $\mathbf{A}_m^a \in \mathbb{C}^{Pk\times Pk}$ is transition matrix, denoting the Kronecker product of identity matrix $\mathbf{I}\in \mathbb{R}^{P\times P}$ and $\mathbf{A}_m^w \in \mathbb{C}^{k\times k}$, and $\mathbf{Q}_m^a \in \mathbb{C}^{Pk\times Pk}$ is measurement matrix, denoting the Kronecker product of $\mathbf{K}^m$'s spatial part $\mathbf{K}^m_{\text{spatial}}=\sum_{r=1}^{R}a_{p}^{m, r}a_{p'}^{m, r}\exp(\ramuno\eta^m\phi_{pp'}^m)$ and $\mathbf{Q}_m^w \in \mathbb{C}^{k\times k}$.

\subsection{Multi-Region Markovian Gaussian Process}

Given our assumption of independence among different dimensions of across-region variables and distinct dimensions of within-region variables, the Markovian representation for all variables $x \in \mathbb{R}^{MP \times T}$, both across- and within-region, spanning $P$ brain regions can be expressed as: \begin{align}\label{all}\begin{split}
f_{t}=\mathbf{A}f_{t-1}+q_{t-1}&, \quad q_{t-1} \sim \mathcal{CN}(0, \mathbf{Q}), 
\end{split}
\end{align} where $f_t=[g_t, \frac{dg_t}{dt}, \dots, \frac{d^{k-1}g_t}{dt^{k-1}}]^{\top} \in  \mathbb{C}^{MPk\times T}$, $x_t$ is the real part of $g_t$, and $\mathbf{A} \in \mathbb{C}^{MPk\times MPk}$, $\mathbf{Q} \in \mathbb{C}^{MPk\times MPk}$ are block diagonal matrices: $\mathbf{A}=\text{diag}\{\mathbf{A}_1^a, \dots, \mathbf{A}_{m_a}^a, \mathbf{A}_1^w \dots \mathbf{A}_{m_w}^w\}$, $\mathbf{Q}=\text{diag}\{\mathbf{Q}_1^a, \dots, \mathbf{Q}_{m_a}^a, \mathbf{Q}_1^w, \dots, \mathbf{Q}_{m_w}^w \}$. Meanwhile, the neural recordings $y\in \mathbb{R}^{N\times T}$ can be reconstructed by $y=\mathbf{C}\operatorname{Re}[\mathbf{G}f]+d+\epsilon$, with $\mathbf{C}$, $d$, $\epsilon$ from CSM-GPFA in Section~\ref{sec:2.2}, and $\mathbf{G}$ in Eq.~\ref{fclds}.

\subsection{Multi-Region Markovian Gaussian Process with Switching States}

After the link between the multi-region Gaussian Process and linear dynamical system (LDS) is established, we can seamlessly extend the  across-region discrete-time LDS in Eq.~\ref{dlds2} to incorporate switching states. 

Integrating a Hidden Markov Model (HMM) into LDS leads to Switching LDS \cite{fox2008nonparametric}, and similarly, combining HMM with MRM-GP results in Switching MRM-GP. A significant advantage of this integration is the ability to link the across-region's transition and measurement matrices with distinct, discrete states $z \in \{1,\dots, Z\}$: $\mathbf{A}_{z}^a=\text{diag}\{\mathbf{A}_{1,z}^a, \dots, \mathbf{A}_{m_a,z}^a\}, \mathbf{Q}_{z}^a=\text{diag}\{\mathbf{Q}_{1,z}^a, \dots, \mathbf{Q}_{m_a,z}^a\}$, which makes it easy to accommodate time-varying frequencies and delays in across-region latent variables.


\section{Inference}

We have now established a connection between a Gaussian Process with a multi-region kernel and a linear dynamical system (LDS). The next step is to learn discrete states, model parameters, and latent variables.

MRM-GP, as a discrete-time LDS, affords a significant advantage: the ability to learn its parameters with a cost linear in time steps: $\mathcal{O}(T)$. To achieve this, we employ the variational Laplace EM inference algorithm proposed in the general recurrent state space framework for decision-making \cite{zoltowski2020general}.

If denoting the number of discrete states as $Z$, the parameters $\theta$ of MRM-GP can be categorized into two groups: (1) kernel parameters: $\{\sigma^{m,z}, \eta^{m, z}\}_{m=1,z=1}^{m_a,Z}$, $\{\sigma^{m,p}, \eta^{m,p}\}_{m=1,p=1}^{m_w,P}$, $\{\phi_{pp'}^{m,z}\}_{m=1, p=1,p'=p+1,z=1}^{m_a, P, Z}$; (2) emissions parameters: $\mathbf{C}, d, \mathbf{V}$. Additionally, the hyper-parameters consist of the number of discrete states $Z$, the number of derivatives $k$, the kernel rank $R$, and the number of latent dimensions $M$. The value of $Z$ depends on the data, $k$ is discussed in Section~\ref{sec:3.1} and Figure~\ref{fig:2}, $M$ is determined through a cross-validation strategy (Section~\ref{experiment:lfp}), and the rank $R$ is consistently set to 2 to ensure positive definiteness without introducing many amplitude parameters. Besides, there is no need to learn the amplitude parameters, denoted as $\{a_{p}^{m,r}\}_{p=1,m=1,r=1}^{P, M, R}$, since the emissions parameter $\mathbf{C}$ fulfills a similar role in MRM-GP.



The variational Laplace EM inference algorithm alternatively updates discrete switching states $z \in \{1,\dots, Z\}$, latent dynamics $f \in \mathbb{C}^{MPk \times T}$, and model parameters $\theta$.  The time complexity and memory storage of each step are all linear in time as follows: (1) updating $z$: $\mathcal{O}(Z)$, $\mathcal{O}(ZT)$; (2) updating $f$: $\mathcal{O}(T)$, $\mathcal{O}(2M^2P^2k^2T)$; (3) updating $\theta$: $\mathcal{O}(ZMk)$, $\mathcal{O}(MPkT)$.

Furthermore, to avoid the calculation of the complex number when updating $f$ in our implementations, we rewrite the complex latent dynamics $f$ in Eq.~\ref{all} to be a joint signal in the real domain, such that the latent dynamics becomes: \begin{align}\label{joint}\begin{split}
&\begin{bmatrix}
    f_r \\
    f_i
\end{bmatrix}_t=\begin{bmatrix}
    \mathbf{A}_r & -\mathbf{A}_i \\
    \mathbf{A}_i & \mathbf{A}_r
\end{bmatrix}\begin{bmatrix}
    f_r \\
    f_i
\end{bmatrix}_{t-1} +\begin{bmatrix}
    q_r \\
    q_i
\end{bmatrix}_{t-1}, \\
&\begin{bmatrix}
    q_r \\
    q_i
\end{bmatrix}_{t-1}\sim \mathcal{N}\rbr{\begin{bmatrix}
    0 \\
    0
\end{bmatrix}, \begin{bmatrix}
    \mathbf{Q}_r & -\mathbf{Q}_i \\
    \mathbf{Q}_i & \mathbf{Q}_r
\end{bmatrix}},
\end{split}
\end{align} where $f_r, f_i, q_r, q_i, \mathbf{A}_r, \mathbf{A}_i, \mathbf{Q}_r, \mathbf{Q}_i$ are the real and imagery part of $f, q, \mathbf{A}, \mathbf{Q}$, respectively.



\section{Experiments}

Our code is available at \url{https://github.com/WeihanLikk/MRM-GP}.

\paragraph{Datasets.} We evaluate MRM-GP on three datasets:\\
$\bullet\quad$ \textbf{Synthetic Data}: We generate simulated data incorporating both across-region communications and within-region neural activities, along with time-varying frequencies and phase delays introduced by various states.\\
$\bullet\quad$ \textbf{Local Field Potential Recordings (LFP)} \cite{siegle2021survey}: Local Field Potential recordings from mouse's primary visual area (V1) and visual anteromedial area (VISam). The external stimulus consisted of an 8Hz drifting grating with eight orientation directions. \\
$\bullet\quad$ \textbf{Neural Spike Trains} \cite{semedo2019cortical, zandvakili2019simultaneous}: Simultaneous spike trains from monkey's primary visual area (V1) and secondary visual cortex (V2). The external stimulus is a 6Hz drifting grating with eight orientation directions.

\paragraph{Baselines for comparison.} We compare MRM-GP with two methods designed to discover the directional communications in the latent space of multi-region recordings: \\
$\bullet\quad$ \textbf{DLAG} \cite{gokcen2022disentangling}: A Gaussian Process Factor Analysis employs a multi-output Squared Exponential kernel. Its goal is to uncover simultaneous or bidirectional latent communications across different regions. The kernel function incorporates a time delay parameter to determine the directions for learned communications.\\
$\bullet\quad$ \textbf{CSM-GPFA}: The Gaussian Process Factor Analysis, using a multi-region kernel as described in Section~\ref{sec:2.2}, is an extension of the model presented in \cite{hultman2018brain}. This extension introduces a new classification assumption, distinguishing latent variables into across-region and within-region types.

\paragraph{Metrics.} For every model and dataset, we fit the model on the training set, denoted as $y_{\text{train}}$, and test its performance on the test set $y_{\text{test}}$. Specifically, we randomly select some trials as $y_{\text{train}}$, while the remaining trials serve as $y_{\text{test}}$. Additionally, we randomly divide the test data $y_{\text{test}}$ into two parts: $y_{\text{test}}^{\text{held-in}}$ with $90\%$ neurons as held-in test data and $y_{\text{test}}^{\text{held-out}}$ with $10\%$ neurons as held-out test data. We infer $x_{\text{test}}^{\text{held-in}}$ based on $y_{\text{test}}^{\text{held-in}}$, which is then used as the test latent variables when computing \textbf{test log-likelihood (LL)} $p(y_{\text{test}}^{\text{held-out}}|x_{\text{test}}^{\text{held-in}}; \theta)$\cite{pei2021neural}, serving as the final metric in our experiments. To reduce the randomness when creating $y_{\text{test}}^{\text{held-in}}$ and $y_{\text{test}}^{\text{held-out}}$, we also average $p(y_{\text{test}}^{\text{held-out}}|x_{\text{test}}^{\text{held-in}}; \theta)$ over five distinct partitions.



\begin{figure}[t]
  \centering
  \includegraphics[width=\linewidth]{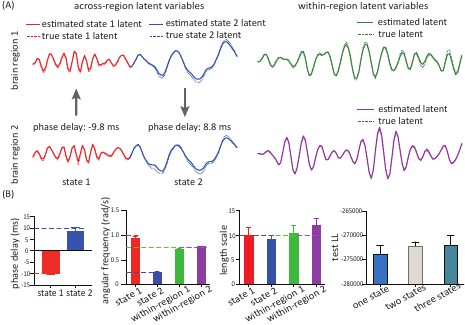}
  \caption{Applying MRM-GP to synthetic data. \textbf{(A)}: Compare the estimated latent variables and discrete states with the ground truth. MRM-GP accurately identifies two states with time-varying frequencies and phase delays, aligning with the ground truth. \textbf{(B)}: Compare learned parameters with the ground truth indicated by dashed lines. \textbf{(C)}: Examine the test LL with varying numbers of discrete states $Z$. The findings indicate that $Z=2$ and $Z=3$ exhibit similar LL, both larger than $Z=1$.}
  \label{fig:3}
\end{figure}

\subsection{Synthetic Data}\label{experiment:synthetic}

This section aims to assess how well MRM-GP can identify switching states, latent variables, and parameters. 

\paragraph{Experimental setup.} We generate 50 independent trials for two brain regions $P=2$, where each region has $30$ neurons, $m_a=1$ dimension across-region variables, and $m_w=1$ dimension within-region variable. We also introduce the time-varying across-region frequencies and phase delays by two discrete states $Z=\{z_1,z_2\}$: (1) state 1: $\eta^{z_1,a}=1.0$ rad/s, $\phi_{1,2}^{z_1}=-10$ ms, $\sigma^{z_1,a}=10$, state 2: $\eta^{z_2,a}=0.25$ rad/s, $\phi_{1,2}^{z_2}=10$ ms, $\sigma^{z_2,a}=10$. Different sign of $\phi_{1,2}$ means the change of directions. We set $\eta^w=0.75$ rad/s and $\sigma^w=10$ for within-region variables. For the generative and inference process, we set hyperparameters $k=2$, $R=2$, and compare the test log likelihood when $Z=1, 2, 3$.


\paragraph{Results.} We fit an MRM-GP to the synthetic data, specifying $m_a=1$ dimension across-region variables, $m_w=1$ dimension within-region variable, and $Z=2$ states. Figure~\ref{fig:3}(A) shows single-trial latent variable estimations that accurately reflect the latent dynamics and communications influenced by discrete states over two brain regions. State $z_1$ (depicted in blue) exhibits a periodic signal with a higher frequency and forward communication from brain region 1 to brain region 2. In contrast, state $z_2$ (shown in purple) displays an oscillatory pattern with a lower frequency and feedback communication from region 2 to region 1.

For a quantitative assessment of learned parameters, Figure~\ref{fig:3}(B) displays the estimated phase delays, frequencies, and length scales across different initializations, demonstrating close alignment with ground truths. Furthermore, Figure~\ref{fig:3}(C) illustrates the test log-likelihood for varying $Z$, revealing that both $Z=2$ and $Z=3$ provide similar and superior estimations compared to $Z=1$ (see Appendix~\ref{syn_z_3} for $Z=3$ visualization). We also have synthetic experiments about parameter initialization and different parameter setting in Appendix~\ref{syn_a1},\ref{syn_a2},\ref{syn_a3}.

\subsection{Local Field Potential Recordings}\label{experiment:lfp}

This section aims to explore interactions between the mouse's primary visual area (V1) and the visual anteromedial area (VISam) in the presence of an 8Hz drifting grating. We also aim to compare the performance and inference time cost with other multi-region methods, namely DLAG and CSM-GPFA.

\paragraph{Experimental setup.} We conduct experiments using two sessions, each comprising eight orientation directions, resulting in 16 datasets. Each dataset includes 15 trials (10 as a training set and 5 as a testing set) of continuous-time local field potential recordings from approximately 20 neurons in V1 and approximately 25 in VISam. The initial sampling rate is 1000Hz, and we downsample it to 100Hz, resulting in 200 time points with 10ms bin size. We set hyperparameters $k=2$, $R=2$.

To determine the dimensionalities of across- and within-region latent variables, we adopt the approach outlined in \cite{gokcen2022disentangling}. Initially, we apply Factor Analysis to identify the total number of latent variables required to elucidate the neural recordings for each region. A 5-fold cross-validation was employed to select the configuration yielding the highest test LL. Subsequently, given the selected total number of latent variables ($M$), we conduct a grid search for the dimensionalities of across- ($m_a$) and within-region ($m_w$), respectively. For each pair of $(m_a, m_w)$, we run 5-fold cross-validation with MRM-GP and chose the setting with the highest test LL. Given this procedure, our final choice was $m_a=1$ across-region variables and $m_w=3$ within-region variables for both V1 and VISam. See Appendix~\ref{sec:apc} for the full comparison.


\begin{figure}[t]
  \centering
  \includegraphics[width=\linewidth]{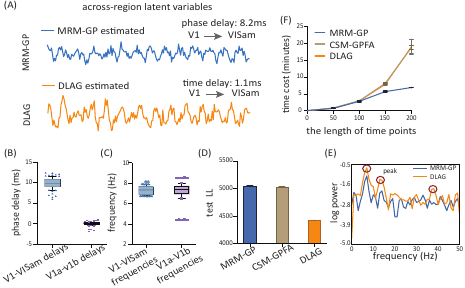}
  \caption{Applying MRM-GP to LFP recordings. \textbf{A}: Compare the across-region latent variables with DLAG. \textbf{B-C}: Visualize the learned phase delays and frequencies from V1-VISam and V1a-V1b. \textbf{D}: Compare the test LL with other multi-region methods: DLAG and CSM-GPFA. \textbf{E}: The power spectrum of across-region latent variables in \textbf{A}. The latent variable from DLAG exhibits three frequency peaks, while MRM-GP's latent variable has only one peak. \textbf{F}: Inference time comparison: MRM-GP has a linear time cost.}
  \label{fig:4}
\end{figure}

\paragraph{Results.} We applied an MRM-GP to local field potential recordings with $Z=1$ state. Figure~\ref{fig:4}(A) shows a comparison of single-trial across-region latent variables for one dataset (orientation 135\degree, session 721123822), and the within-region variables are in Appendix~\ref{sec:apc}. Both latent variables demonstrate an oscillatory structure, capturing characteristics of the external 8Hz drifting grating stimulus. 


The MRM-GP's latent variable in this dataset is linked to a communication direction from V1 to VISam with an 8.2 ms phase delay, and the DLAG's latent variable shows a 1.1 ms time delay. Both delays fall within a single time bin (10ms) and are positive, suggesting consistent communication direction from V1 to VISam. The difference in their values arises because DLAG models time delay ($\delta_{pp'}$) in the kernel equation $K_{pp'}(\tau)=\exp(-\frac{1}{2\sigma^2}(\tau-\delta_{pp'})^2)$, which is independent of frequency and has a different interpretation from the phase delay ($\phi_{pp'}$). In this context, $\phi_{pp'}$ represents the delay in a specific frequency band. In contrast, $\delta_{pp'}$ signifies the delay for a latent variable with a mixture of multiple frequencies, as evidenced by its power spectrum with three frequency peaks (Figure~\ref{fig:4}(E)). Therefore, the divergence in values is acceptable if their directions align.

The left chart in Figure~\ref{fig:4}(B) illustrates the estimated phase delays across all 16 datasets. Each data point represents the phase delay for an individual run on a specific dataset. The findings suggest consistent communication from V1 to VISam across all datasets, with the phase delays clustered around 7.5Hz (the left chart in Figure~\ref{fig:4}(C)), which is consistent with external 8Hz stimulus. 

To demonstrate that the MRM-GP itself is not the cause of delays, we first divide V1 randomly into two parts, V1a and V1b, each with channels of equal size. We then estimated the phase delays between them. The right chart in Figure \ref{fig:4}(B) illustrates that across 16 datasets, all phase delays hover around zero within frequency bands around 7.5Hz (the right chart in Figure \ref{fig:4}(C)). This suggests that the learned delays are a consequence of the data rather than the model. 

Figure~\ref{fig:4}(D) shows that MRM-GP, a linear dynamics system approximation of CSM-GPFA, exhibits a similar test LL to CSM-GPFA. The higher test LL compared to DLAG suggests that the multi-region kernel (Eq.~\ref{csm}) outperforms DLAG's Squared Exponential kernel on these datasets. This is attributed to the former explicitly modeling frequencies through its kernel parameters and having a better frequency separation. Specifically, the across-region variable of the former has only one prominent frequency, whereas DLAG's across-region variable exhibits three peaks in Figure\ref{fig:4}(E), keeping consistent with the data spectrum in Appendix~\ref{data_spectrum}.

Lastly, in Figure \ref{fig:4}(F), we compare the inference time of MRM-GP, CSM-GPFA, and DLAG for 500 iterations. We achieved this by downsampling the recordings and creating four datasets with varying lengths of time points: 50, 100, 150, and 200. The results indicate that the time cost of MRM-GP increases linearly, whereas both CSM-GPFA and DLAG exhibit cubic growth.


\subsection{Neural Spike Trains}\label{experiment:spike}

This section aims to evaluate MRM-GP's ability to identify switching states within the communications subspace while also discovering across-region communications using a distinct type of neural data.

\paragraph{Experimental setup.} The simultaneous spike trains were obtained from the monkey's primary visual area (V1) and secondary visual cortex (V2) in the presence of a 6Hz moving grating. This dataset comprises four sessions, each featuring eight orientation directions, resulting in 32 datasets. Each dataset comprises 400 trials (64 time points with 20ms bin size for every trial), with 300 trials randomly selected as the training set and 100 trials as the testing set. In V1, there are approximately 90 neurons, while in V2, there are around 20 neurons. We set hyperparameters $k=2$, $R=2$.

\paragraph{Results.} We fitted an MRM-GP to neural spikes trains with $m_a=2$ dimensions across-region variables, $m_w=2$ dimension within-region variable, and $Z=2$ states. The configuration of $m_a$ and $m_w$ follows previous work \cite{gokcen2022disentangling} and adopts the same strategy mentioned in Section~\ref{experiment:lfp}. 

\begin{figure}[t]
  \centering
   \includegraphics[width=\linewidth]{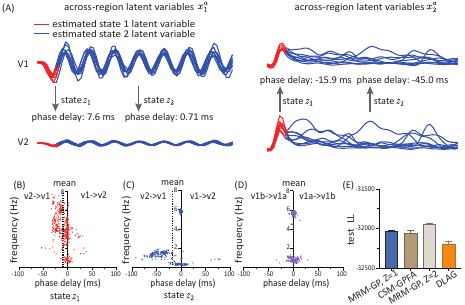}
  \caption{Applying MRM-GP to neural spike trains. \textbf{A}: Visualize across-region latent variables with state $z_1$ (in red) and state $z_2$ (in blue). The results suggest forward and feedback communications between V1 and V2, along with time-varying delays. \textbf{B-C}: Visualize estimated parameters through phase delays on the x-axis and frequencies on the y-axis. \textbf{D}: Depict phase delays and frequencies learned in the V1a-V1b control experiment, with their representation on the x-axis and y-axis, respectively. \textbf{E}: Compare the test LL with discrete states $Z=1,2$ and other multi-region methods: DLAG and CSM-GPFA.}
  \label{fig:5}
\end{figure}

Figure~\ref{fig:5}(A) shows the across-region latent variables for one dataset (orientation 0\degree, session 106r001p26, ten trials are displayed, all variables are scaled by the variance explained in each region), and the within-region variables are in Appendix~\ref{sec:apc}. These latent variables indicate time-varying forward and feedback communications between V1 and V2. Different states exhibit distinct phase delays and frequencies. The first dimension of across-region variables (denoted as $x_1^a$) displays a periodic pattern caused by the external drifting grating stimulus, whereas the second dimension (denoted as $x_2^a$) exhibits a non-periodic signal with a single peak shortly after the stimulus onset.

Figure~\ref{fig:5}(B-C) presents the estimated phase delays and frequencies over multiple independent runs for 32 datasets. Each data point represents a dimension of across-region variables. Figure~\ref{fig:5}(B) corresponds to state $z_1$, indicating that most state $z_1$ dimensions exhibit across-region interactions within the 2Hz-8Hz range. Additionally, some dimensions display feedback communication with a large phase delay ($>$10ms) from V2 to V1, corresponding to state $z_1$ of  $x_2^a$ in Figure~\ref{fig:5}(A). However, there is variability across datasets for state $z_1$ with smaller phase delays ($<$10ms). Some show forward communication from V1 to V2, akin to state $z_1$ of $x_1^a$ in Figure~\ref{fig:5}(A), while others indicate feedback communications from V2 to V1.

One explanation for this variability is that in certain datasets, $x_1^a$ has a much weaker amplitude compared to $x_2^a$, making the weaker latent affected by the stronger one along with its delay. This leads to a feedback signal from V2 to V1 at state $z_1$. On the other hand, in some datasets (e.g., orientation 0\degree, session 106r001p26), $x_1^a$ is not as weak, resulting in a forward signal from V1 to V2 at state $z_1$.

Figure~\ref{fig:5}(C) depicts the estimated phase delays and frequencies associated with state $z_2$. The findings indicate a clear separation of oscillatory communications into two frequencies. One involves 6Hz communications with small phase delays (referring to state $z_2$ of $x_1^a$ in Figure~\ref{fig:5}(A)), while the other involves 1Hz communications with large phase delays (akin to state $z_2$ of $x_2^a$ in Figure~\ref{fig:5}(A)).

The time-varying phase delays can be explained as follows: (1) For $x_1^a$, V1 triggers V2 to have oscillatory dynamics during state $z_1$, while in state $z_2$, V2 is already engaged, causing both regions to oscillate synchronously, resulting in a smaller phase delay than in state $z_1$. (2) For $x_2^a$, V2 consistently sends signals with a low frequency to V1, resulting in a larger phase delay due to the longer period as indicated by $z_2$. During state $z_1$, the stimulus onset triggers an intense signal from V2 to V1, leading to a smaller phase delay, which can be considered as an emergence of surprise or prediction error from V2 to V1 \cite{rao1999predictive}.

Similar to Section~\ref{experiment:lfp}, we also perform a control experiment by learning the phase delays between V1a and V1b. In Figure~\ref{fig:5}(D), the outcomes reveal zero-delay communications that are distributed across two frequencies (6Hz, 1Hz), suggesting that learned delays are a consequence of the data rather than the model. 

Finally, we compare the test LL of MRM-GP, CSM-GPFA, and DLAG in Figure~\ref{fig:5}(E). The results indicate that MRM-GP with $Z=2$ states achieves the highest LL, while MRM-GP with $Z=1$ state exhibits a similar LL compared to CSM-GPFA, and both outperform DLAG. This suggests that (1) switching states exist in these datasets; (2) the multi-region kernel is more appropriate than the Squared Exponential kernel for modeling signals with sinusoidal structures.


\section{Discussion}

MRM-GP establishes the connection between a linear dynamics system (LDS) and a multi-output Gaussian Process (GP) explicitly modeling frequency-based communications and their directionality via phase delays within the latent space of neural data.

Connecting a complex-valued GP with an LDS is non-trivial. Although a complex-valued GP can be written as a multi-output real-valued GP (by twice the dimension), the resulted multi-output GP cannot be converted to an LDS by spectral factorization because the separability of the resulted multi-output kernel is not guaranteed.

Once the link is established, we can harness several advantages: (1)~using the powerful representational
capability of kernels, such as applying a multi-region kernel to model latent variables with periodic patterns; (2)~achieving a linear computational cost; (3)~incorporating time-varying frequencies and delays by introducing different discrete states.

We test MRM-GP using two distinct types of neural data. The findings showcase its capability to discover state-dependent latent communications across brain regions with a linear time inference cost.

Finally, the limitations of MRM-GP are twofold: (1) its reliance on separability for multi-output kernels, which restricts kernel selection options, and (2) its current model assumptions are unable to capture phenomena such as phase resetting and phase variability across different trials.

\section*{Acknowledgement}

This work is supported by National Institutes of Health BRAIN initiative (1U01NS131810).

\section*{Impact Statement}

The MRM-GP introduces an innovative and efficient method for investigating intricate interactions among brain regions. Its capacity to deliver an interpretable representation of multi-region neural data is poised to advance neuroscience, offering the potential for a more profound comprehension of brain function and disorders. This enhanced understanding of brain interactions has the prospect to drive advancements in neurotechnology, with potential benefits extending to fields such as brain-computer interfaces and personalized medicine.

\nocite{langley00}

\bibliography{main}
\bibliographystyle{icml2024}

\newpage
\appendix
\onecolumn

\section{Spectral Factorization.}\label{sec:apb}

\paragraph{Derivation for Eq.~\ref{fclds}.}

Start with Eq.~\ref{clds}, taking Fourier transforms on both sides gives:\begin{align}\label{supp1}\begin{split}
(\ramuno\omega) \mathbf{J}(\ramuno\omega)^{p, w}_m&=\mathbf{F}_m^w \mathbf{J}(\ramuno\omega)^{p, w}_m+\mathbf{L} \mathbf{U}(\ramuno\omega).
\end{split}
\end{align} Solving for $ \mathbf{J}(\ramuno\omega)^{p, w}_m$ gives:\begin{align}\label{supp2}\begin{split}
 \mathbf{J}(\ramuno\omega)^{p, w}_m=((\ramuno\omega-\mathbf{F}_m^w)\mathbf{I})^{-1}\mathbf{L} \mathbf{U}(\ramuno\omega).
\end{split}
\end{align}

Recall that $f(t)^{p, w}_m$ in Eq.~\ref{clds} is a Complex Gaussian Process with single-output kernel $\mathbf{K}(\tau)^m=\sum_{r=1}^{R}{a_{r}^{m}}^2\exp(-\frac{1}{2{\sigma^m}^2}\tau^2+\ramuno\eta^m(\tau))$ and its derivatives up to $(k-1)^{th}$. So, the spectral density matrix of this process $f(t)^{p, w}_m$ is: \begin{align}\label{supp3}\begin{split}
 \mathbf{S}_J(\omega)=\mathbb{E}[ \mathbf{J}(\ramuno\omega)^{p, w}_m{J(-\ramuno\omega)^{p, w}_m}^{\top}].
\end{split}
\end{align}

Bring Eq.~\ref{supp2} into Eq.~\ref{supp3} gives: \begin{align}\label{supp4}\begin{split}
 \mathbf{S}_J(\omega)=&(\mathbf{F}_m^w-\ramuno\omega\mathbf{I})^{-1}\mathbf{L}\mathbb{E}[ \mathbf{U}(\ramuno\omega)U(-\ramuno\omega)^{\top}]\mathbf{L}^{\top}(\mathbf{F}_m^w+\ramuno\omega\mathbf{I})^{-T},\\
&=(\mathbf{F}_m^w-\ramuno\omega\mathbf{I})^{-1}\mathbf{L}v\mathbf{L}^{\top}(\mathbf{F}_m^w+\ramuno\omega\mathbf{I})^{-T}.
\end{split}
\end{align}

Finally $S(\omega)$ in Eq.~\ref{fclds} is: \begin{align}\label{supp5}\begin{split}
S(\omega) = \mathbf{G} \mathbf{S}_J(\omega)\mathbf{G}^{\top}.
\end{split}
\end{align}

\paragraph{Finding roots for $T(\ramuno\omega)$ in Eq.~\ref{sf}.} Using the $T(\ramuno\omega)$'s coefficients $b_0, b_1, \dots, b_{2k}$, we can create a companion matrix: \begin{equation}\label{companion}\begin{split}
\mathbf{B}&=\begin{bmatrix}
    0 & 0 & \dots & 0 & -\frac{b_{2k}}{b_0}  \\
    1 & 0 & \dots & 0 & -\frac{b_{2k-1}}{b_0}  \\
    \vdots & \vdots & \dots & \vdots & \vdots \\
    0 & 0 & \dots & 1 & -\frac{b_{1}}{b_0}  \\
\end{bmatrix},
\end{split}\end{equation} where the eigenvalues of this matrix are the roots for $T(\ramuno\omega)$ \cite{hom1985matrix}. Notably, the companion matrix is structured as a Hessenberg matrix, suggesting that its eigenvalues can be obtained through the QR algorithm with Givens rotation. This process has a time complexity of $\mathcal{O}(k)$ for each iteration.

\clearpage
\section{DLAG latent variables for V1-V2 spike train data}\label{sec:apc}

\begin{figure}[!ht]
  \centering
  \includegraphics[width=\linewidth]{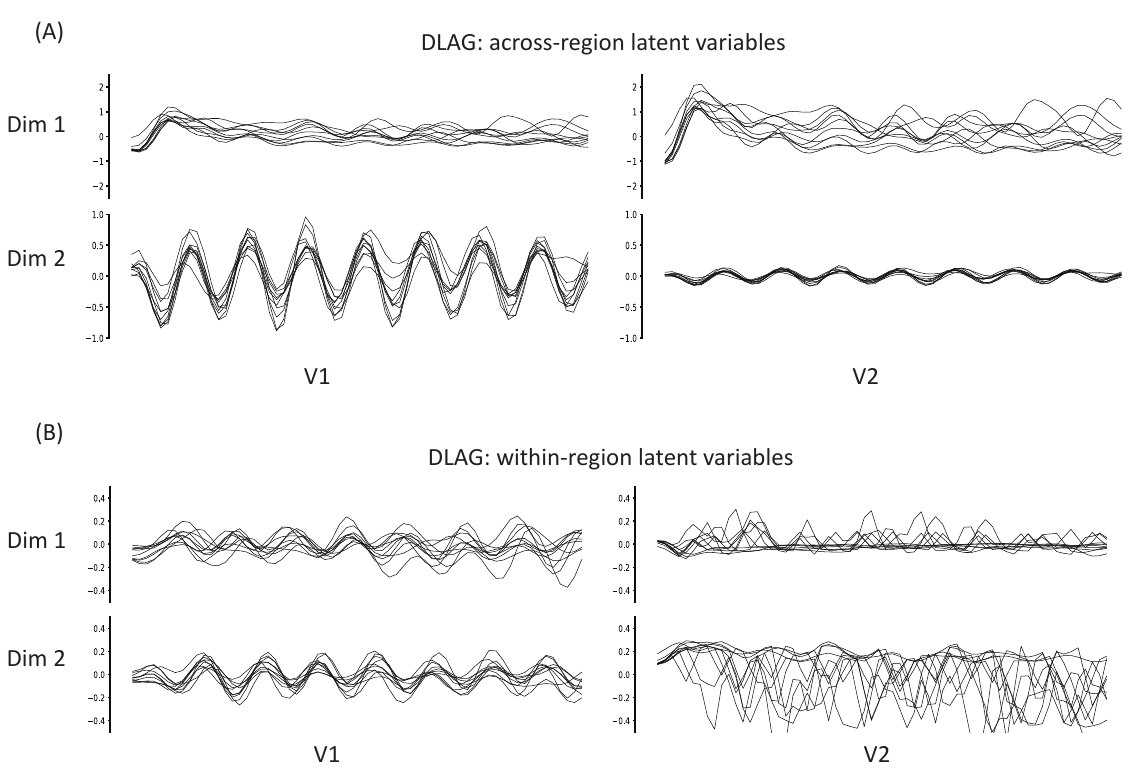}
  \caption{Across- and within-region latent variables for DLAG on V1-V2 spike train data. DLAG's two across-region latent variables both have some oscillations. However, there are fewer oscillations in the first dimension of our model's across-region variables (Figure 5), indicating a better disentanglement of frequencies.}
  \label{fig:re1}
\end{figure}

\clearpage
\section{Within-region latent variables for V1-V2 spike train data}
\begin{figure}[!ht]
  \centering
  \includegraphics[width=\linewidth]{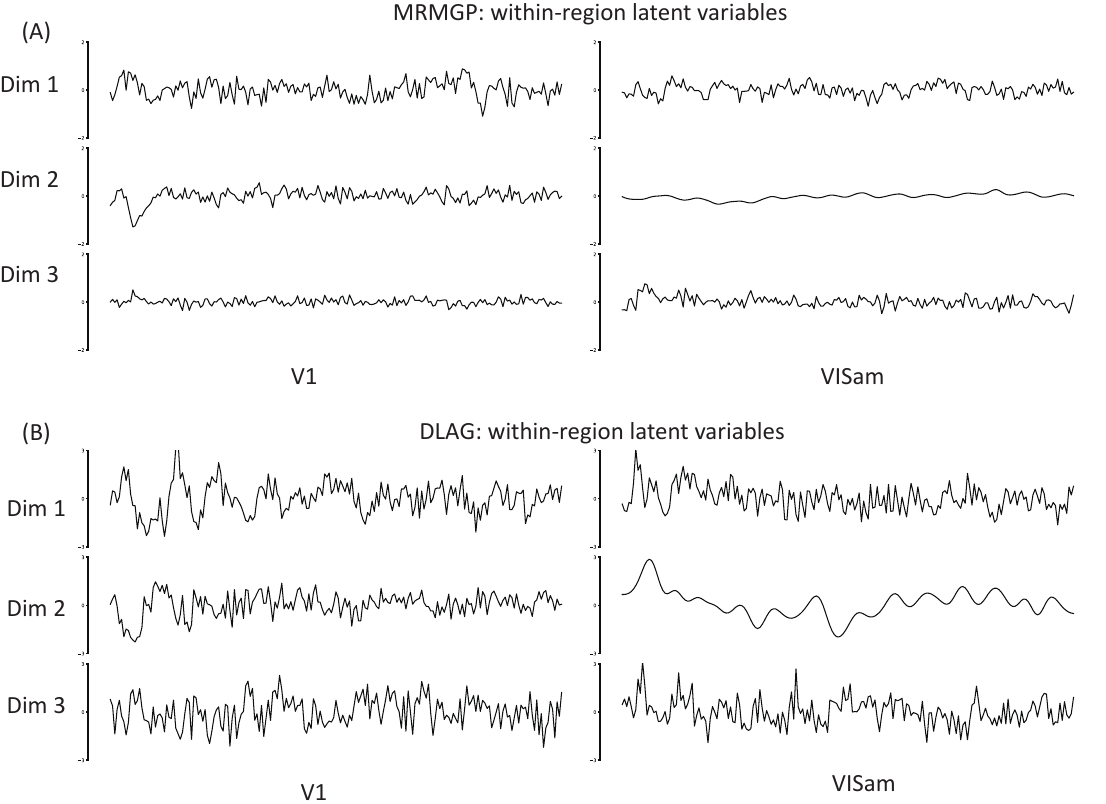}
  \caption{Within-region latent variables for MRM-GP and DLAG on local field potential recordings from V1 and VISam. MRMGP and DLAG exhibit similar within-region latent variables associated with high-frequency noise, indicating neural activities specific to a certain region.}
  \label{fig:supp2}
\end{figure}

\clearpage
\section{Compare latent variables with different k}\label{com_k}

\begin{figure}[!ht]
  \centering
  \includegraphics[width=\linewidth]{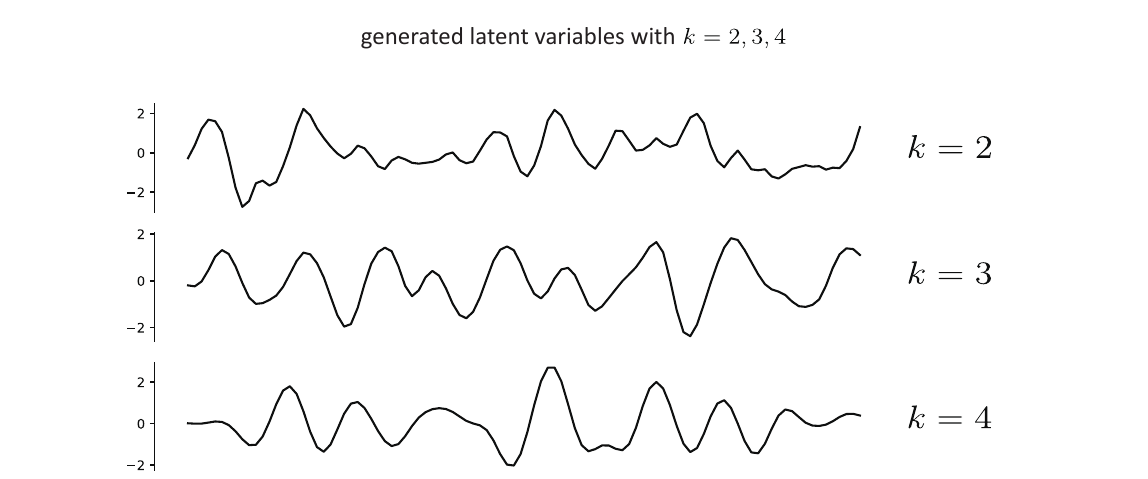}
  \caption{Compare the smoothness of latent variables with $k=2,3,4$. We generate latent variables with the same frequency, length scale and phase delays, while changing the value of $k$. The results show that a larger $k$ makes the latent variable smoother.}
  \label{fig:re2}
\end{figure}

\clearpage
\section{Spectrum of the LFP data}\label{data_spectrum}

\begin{figure}[!ht]
  \centering
  \includegraphics[width=\linewidth]{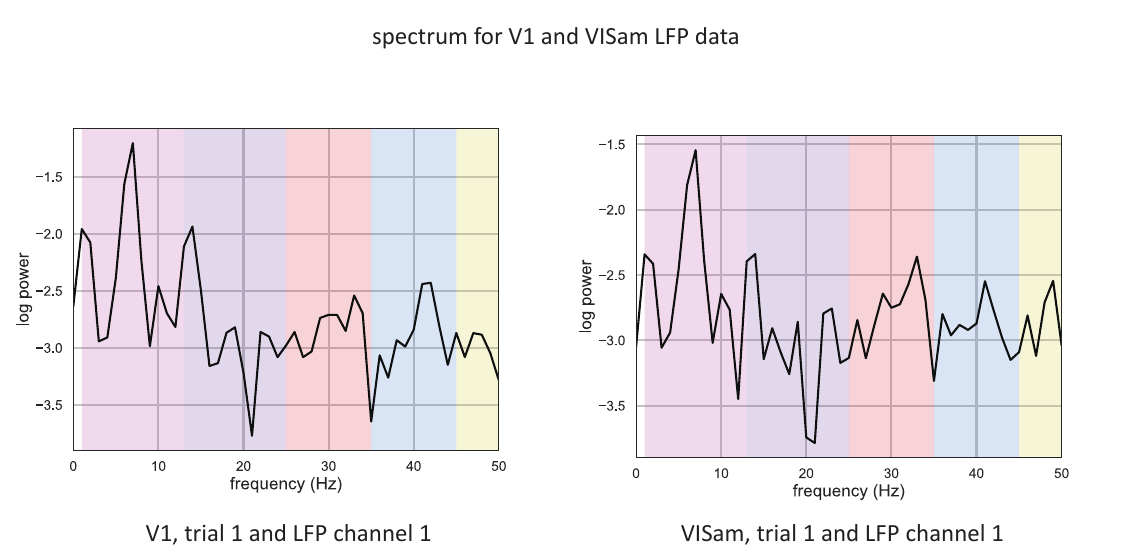}
  \caption{The spectrum for V1 and VISam LFP data with one trial and one channel shown. The multiple peaks in both V1 and VISam spectra indicate a mix of frequencies. In Figure 4, the spectrum of DLAG's across-region latent variable has three peaks, while our model's across-region latent variable has only one dominant peak, indicating a better frequency separation brought by the multi-region kernel in Eq.1.}
  \label{fig:re2}
\end{figure}

\clearpage
\section{Synthetic data with a wide range of frequencies and delays}\label{syn_a1}

\begin{figure}[!ht]
  \centering
  \includegraphics[width=\linewidth]{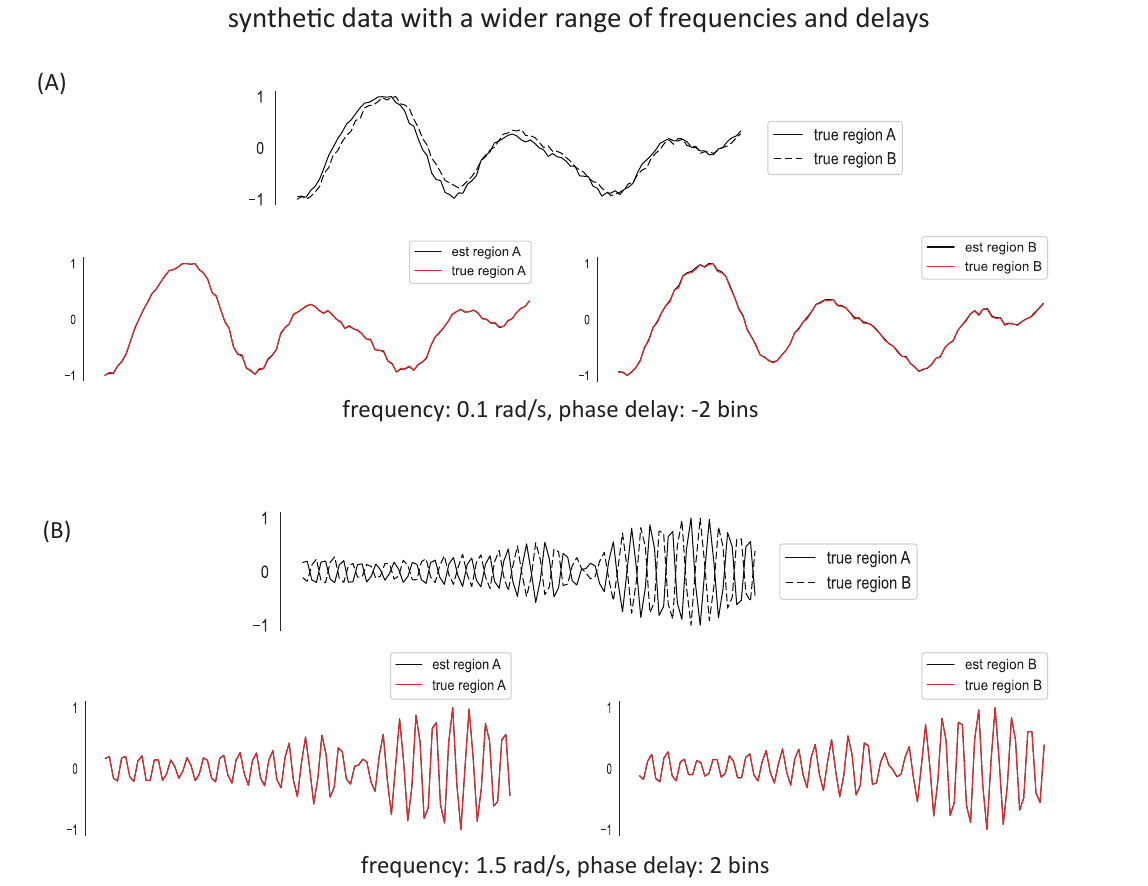}
  \caption{Synthetic data experiments with a wide range of frequencies and delays. We tested our model on a wide range of parameters, where the frequencies vary from 0.1 rad/s to 1.5 rad/s, and the delays vary from -2 to 2. Under all these cases, our model could perfectly recover the latent variables.}
  \label{fig:re2}
\end{figure}

\clearpage
\section{Synthetic data with different parameter initialization}\label{syn_a2}

\begin{figure}[!ht]
  \centering
  \includegraphics[width=\linewidth]{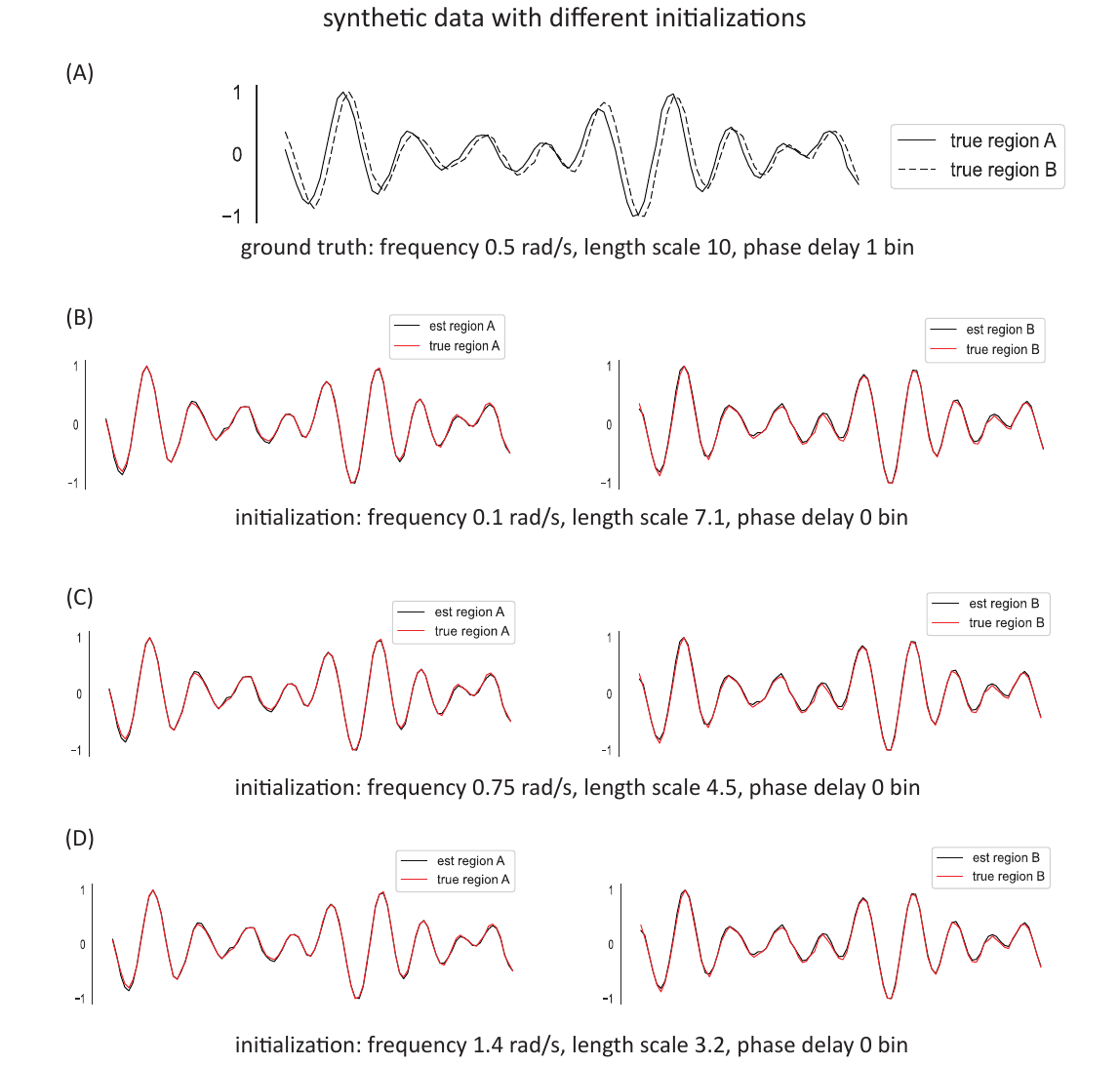}
  \caption{Synthetic data experiments with different parameter initialization. We tested our model with three initialization settings: from smoothness to unsmoothness. Under all these cases, our model could perfectly recover the latent variables, indicating that our model is not very sensitive to parameter initialization.}
  \label{fig:re2}
\end{figure}

\clearpage
\section{Synthetic data without pure oscillation}\label{syn_a3}

\begin{figure}[!ht]
  \centering
  \includegraphics[width=\linewidth]{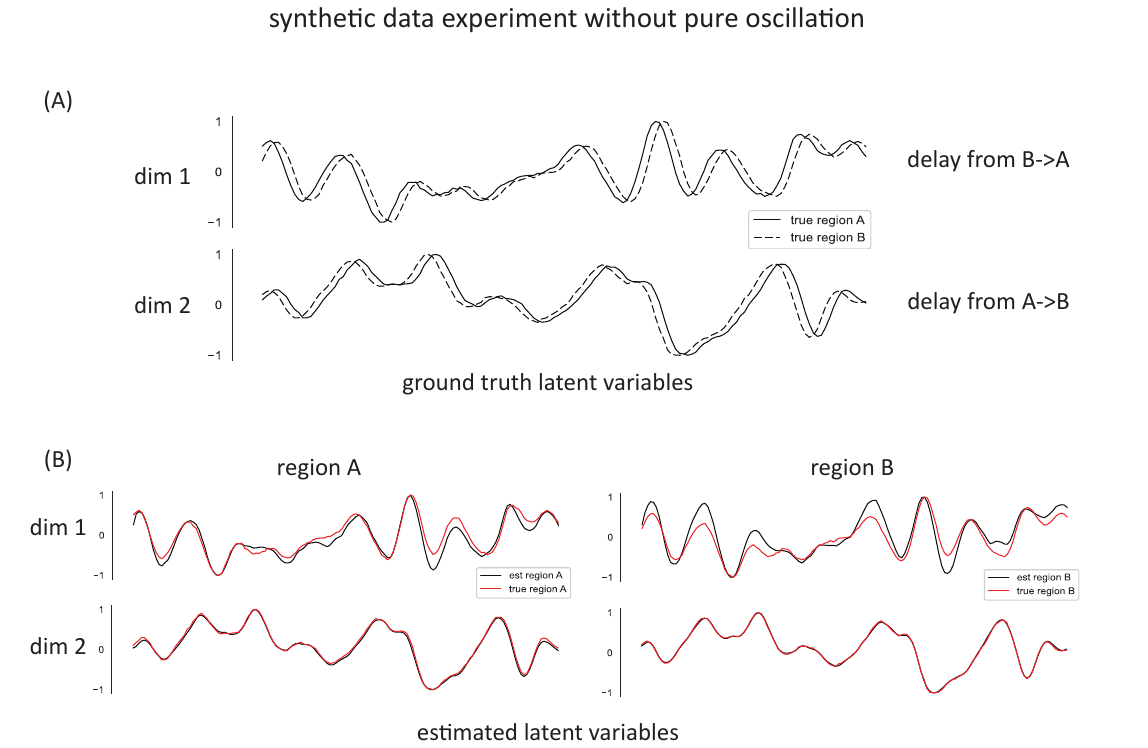}
  \caption{Synthetic data experiment without pure oscillation. We generated latent variables from a Gaussian Process with SE kernel. We added a manual shift to create delays, where one latent communication has a positive delay from Region A to Region B while the other has an opposite delay. The results show that our model can recover the true latent variables and correct delays between different regions. In summary, this demonstrates our model's ability to handle non-periodic data.}
  \label{fig:re2}
\end{figure}

\clearpage
\section{Cross-validating for the dimensionality of MRM-GP}
\begin{figure}[!ht]
  \centering
  \includegraphics[width=\linewidth]{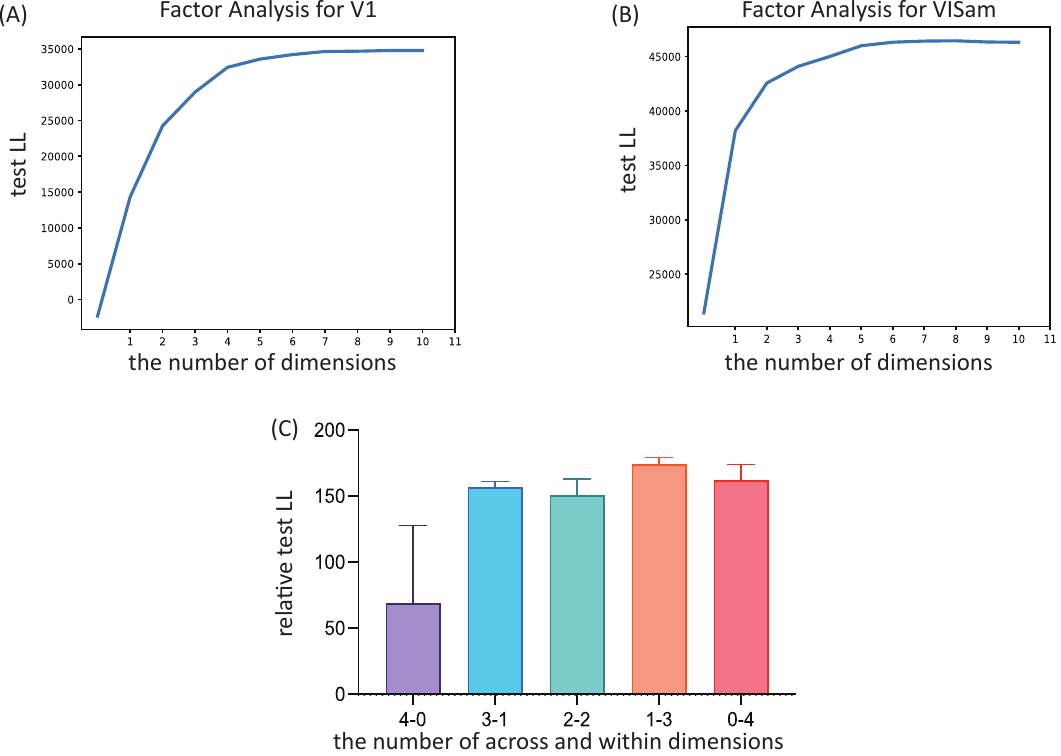}
  \caption{Determine the number of dimensions: (A-B) Factor Analysis is applied to V1 and VISam, revealing a gradual increase in test log-likelihood (LL) when the number of dimensions surpasses four (averaged over 5-fold cross-validation) . Considering both the inference time for all three models, we opt for four as the total dimensions for each region. (C) Conducting a grid search for across- and within-region variables using relative test LL as the metric—the difference between the model's test LL and Factor Analysis's test LL. The outcomes indicate that one across-region and three within-region variables yield the highest LL.}
  \label{fig:supp1}
\end{figure}



\clearpage
\section{Applying MRM-GP to synthetic data with $Z=3$}\label{syn_z_3}

\begin{figure}[!ht]
  \centering
  \includegraphics[width=\linewidth]{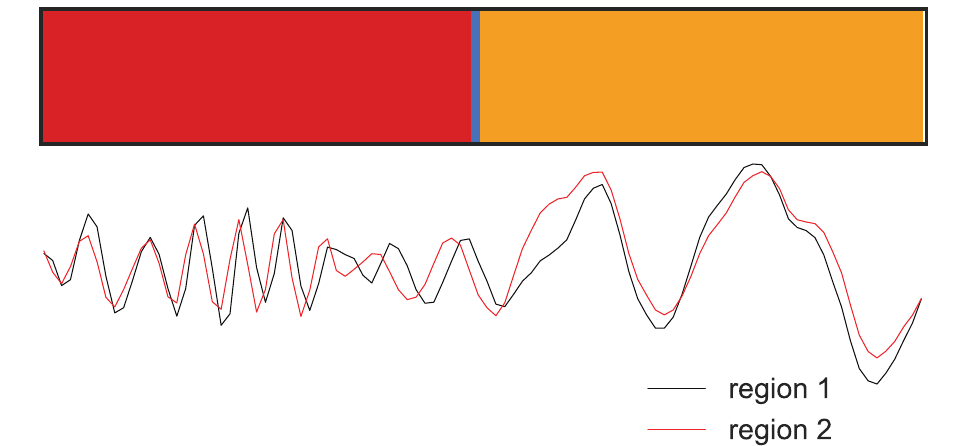}
  \caption{Applying MRM-GP to synthetic data with $Z=3$. The result closely resembles Figure~\ref{fig:3} when $Z=2$, suggesting that the model can effectively learn the correct number of discrete states even specifying a large $Z$.}
  \label{fig:supp4}
\end{figure}

\end{document}